\documentclass[epj]{svjour}
\usepackage{graphics}
\usepackage{commath}
\usepackage{amsmath}
\usepackage{amssymb}
\usepackage{color}
\usepackage{url}
\usepackage{graphicx}
\usepackage{bm}
\usepackage{hyperref}
\usepackage{multicol}
\usepackage{braket}

\hypersetup{colorlinks=true,citecolor=blue, linkcolor=blue, urlcolor=blue}
\begin{document}
\title{Role of a magnetic field in the context of inhomogeneous gravitational collapse}

\author{Shibendu Gupta Choudhury\thanks{\emph{email:} sgc14ip003@iiserkol.ac.in}%
}                     
%
%
\institute{Department of Physical Sciences,  Indian Institute of Science Education and Research Kolkata, Mohanpur, West Bengal 741246, India.}
\date{Received: date / Revised version: date}
%
\abstract{
Magnetic fields have been found to have an inherent capability of acting against gravity. An important question posed in the literature is whether presence of a magnetic field can  alter the dynamics of a gravitational collapse and prevent the final formation of a singularity. Inhomogeneous models of collapse have not been explored significantly in this context. In the present work we  investigate the role of magnetic fields in the evolution of inhomogeneous cylindrically symmetric models. We use an approach based on the Raychaudhuri equation for such an analysis. We show that it is quite possible for the magnetic field to avert the gravitational collapse in these models.
\PACS{
      {PACS-key}{discribing text of that key}   \and
      {PACS-key}{discribing text of that key}
     } 
} 
\maketitle
\section{Introduction}
A massive astronomical body undergoes the phase of a gravitational collapse when the pressure inside it  fails to support the gravitational pull. Gravitational collapse is a widely studied subject which began after the pioneering works of Datt\cite{datt} and Oppenheimer and Snyder\cite{os}. They investigated an idealized star model which collapses to zero proper volume. Later there have been many attempts to explore more general and physically acceptable collapse scenarios with different kinds of matter distribution and content. For an elegant and detailed discussion on the current state of the subject, as well as the open problems, we refer to\cite{joshi1,joshi2}. 

It is generally believed that gravitational collapse leads to the formation of curvature singularities as the end state. It has been proved by Penrose and Hawking\cite{penrose,hawking,hawell} that in General Relativity (GR) singularities are inevitable for purely gravitational systems. All physical laws including GR break down and physical quantities like energy density, pressure, curvature etc. diverge at a singularity. Therefore, finding an escape route from these singularities is needed to have a complete physical description.

 An important question in the current context is whether non-gravitational agents (e.g. electromagnetism) can alter the fate of gravitational collapse so that formation of singularity can be avoided. A significant amount of research has been done in this direction which can be found in \cite{tsagas1,tsagas2,tsagas3,tsagas4,tsagasmr,novi,cruz,rayc1,arda,ori,feli1,feli2,ray,ghez,kras} and references therein. For example, the role of repulsive Coulomb forces in gravitational collapse has been studied in \cite{novi,cruz,rayc1,kras,tsagas3}. In the present work we consider magnetic field as the non-gravitational agent. Considering presence of magnetic fields is known to be a well-motivated choice in this context for several decades\cite{thornecp}. This idea first emerged in the literature with the works of Melvin\cite{melvin1,melvin2}. He proposed 
a cylindrically symmetric non-singular static solution which describes a magnetic universe\cite{melvin1}. In the Melvin spacetime magnetic field gives rise to a pressure which balances the gravitational pull. The stability of this solution was considered in \cite{melvin2,thorne1}. Thorne established that the solution is absolutely stable against radial perturbations and does not collapse to a singularity or undergo dispersal\cite{thorne1}. Following these works, it has been found by many authors that magnetic fields do have an inherent ability to resist the gravitational contraction\cite{tsagas1,tsagas2,tsagas3,tsagas4,arda,tsagas5}. This is because magnetic force lines repel each other. Presence of magnetic field offers a `magneto-curvature stress' which opposes any agent that tries to distort the magnetic field lines\cite{tsagas2,tsagas4,tsagas5,tsagas6}. 
 
 To get a complete picture regarding the dynamics of a gravitational collapse, we need to solve the full set of non-linear Einstein equations. It is actually a very difficult job to find exact solutions for a sufficiently general system. Therefore, simplified models incorporating the essential physics are explored in the context of gravitational collapse. An alternative way to study this subject is to use an approach based on the Raychaudhuri equation\cite{rc,ehlers}. One has to use the idea of considering the flow lines of particles inside the collapsing distribution as the congruence for such a study. The utility of applying the Raychaudhuri equation lies in the fact that one can often draw important conclusions without solving the field equations. We shall see this more explicitly in section \ref{sec3} where we briefly review the homogeneous models. The idea of applying Raychaudhuri's equation in collapsing situations was introduced by Germani and Tsagas in the context of magnetized Tolman-Bondi collapse\cite{tsagas1}. Later this technique has been applied in a few other models of gravitational collapse\cite{tsagas2,tsagas3,tsagas4,tsagasmr,sgc}.

Though there are indications that magnetic fields can provide a way of avoiding formation of singularities in a gravitational collapse, definite conclusions regarding this hardly exist. Therefore, we need further investigation in this direction. In this work, our aim is to study the role of magnetic fields towards deciding the fate of an inhomogeneous gravitational collapse. We use the Raychaudhuri equation and its most important consequence, namely the focusing condition\cite{wald,poisson} for this purpose. Our work is motivated by the ideas proposed in a few recent studies\cite{tsagas2,tsagas3,tsagas4}. In these contributions, the authors worked out a crucial condition  which determines the final fate of a gravitational collapse in the presence of a magnetic field. This condition simply means that singularities do not form if the magneto-curvature stress overcomes the gravitational attraction. They have used a covariant approach\cite{tsagas3,tsagas4} and the corresponding Raychaudhuri equation to arrive at this condition. Later, Mavrogiannis and Tsagas\cite{tsagasmr} examined the possibility that the aforementioned condition is satisfied at some stage during the collapse in an almost homogeneous background, namely perturbed Bianchi I background. For further investigation in this direction, in the present work we consider the evolution of cylindrically symmetric inhomogeneous spacetimes consisting of a magnetic field and a charged fluid which obeys the strong energy condition. We carry out the analysis within the ideal--Magnetohydrodynamics (ideal--MHD) limit\cite{priest}.

The structure of the paper is as follows. In the next section (section \ref{sec2}) we discuss the Raychaudhuri equation and the focusing condition for timelike congruences. 
We use these concepts in the subsequent sections. In section \ref{sec3} we briefly review the case of homogeneous models to emphasize on the utility of using the Raychaudhuri equation. Investigation of the dynamics of inhomogeneous spacetimes using the Raychaudhuri equation is carried out in section \ref{sec4}. We consider two specific cases as examples. In the first one, it is assumed that the spatial and temporal parts of the metric coefficients are separable while the spacetime is assumed to be self-similar in second case. The final section \ref{sec5} contains some concluding remarks.

\section{The Raychaudhuri equation and focusing condition}\label{sec2}

 The Raychaudhuri equation for a timelike congruence is given by\cite{rc,ehlers},
\begin{equation}
\label{raych-eq}
 \frac{\mathrm{d}\theta}{\mathrm{d}\tau}=-\frac{1}{3}\theta^2+\nabla_{\alpha}a^{\alpha}-\sigma_{\alpha\beta}\sigma^{\alpha\beta}
 +\omega_{\alpha\beta}\omega^{\alpha\beta}-R_{\alpha\beta}u^\alpha u^\beta.
 \end{equation}
 Here $u^\alpha$ is the velocity vector of the congruence;
$\theta=\nabla_\alpha u^\alpha$ is the expansion scalar; $\tau$ is proper time; 
$\sigma_{\alpha\beta}=\nabla_{(\beta}u_{\alpha)}-\frac{1}{3}h_{\alpha\beta}\theta+a_{(\beta}u_{\alpha)}$
is the shear tensor where $h_{\alpha\beta}$ is the induced spatial metric; $\omega_{\alpha\beta}=\nabla_{[\beta}u_{\alpha]}-a_{[\beta}u_{\alpha]}$ 
is the rotation tensor; $a^\alpha=u^\beta\nabla_{\beta}u^{\alpha}$ is the acceleration vector and $R_{\alpha\beta}$ is the Ricci tensor.

An initially converging congruence will focus ($\theta\rightarrow -\infty$) within a finite proper time if\cite{wald,poisson}, 
\begin{equation}\label{fc}
 \frac{d\theta}{d\tau}+\frac{1}{3}\theta^2\leq 0.
\end{equation}
This condition is known as the focusing condition. 
For hypersurface orthogonal, geodesic congruences we have $\omega_{\mu\nu}=0$\cite{wald,poisson} and $a^\mu=0$. In GR, the strong energy condition, 
\begin{equation}
 T_{\alpha\beta}u^\alpha u^\beta+\frac{1}{2}T\geq 0 \implies R_{\mu\nu}u^\mu u^\nu\geq 0.
\end{equation}
Here, $T_{\alpha\beta}$ denotes the energy-momentum tensor and $T$ is its trace. The strong energy condition is a physically well-motivated assumption. All known matter distribution are found to obey this condition. The condition $R_{\mu\nu}u^\mu u^\nu\geq 0$ means that gravity is attractive. Now, from the Raychaudhuri equation \eqref{raych-eq} we have,
\begin{equation}\label{rcaf}
  \frac{d\theta}{d\tau}+\frac{1}{3}\theta^2=-\sigma_{\alpha\beta}\sigma^{\alpha\beta}
 -R_{\alpha\beta}u^\alpha u^\beta.
\end{equation}
As $\sigma_{\alpha\beta}$ is purely spatial,  $\sigma_{\alpha\beta}\sigma^{\alpha\beta}\geq 0$. Therefore, the focusing condition is always satisfied for these congruences. 

Focusing within a finite interval cannot be prevented for purely gravitational systems (which do not contain any non-gravitational agent) in GR. This is because all the terms in the right hand side of the equation \eqref{rcaf} are negative which helps in the convergence of the worldlines within the congruence. So, these terms assist gravitational implosion. Therefore, it is expected that gravitational collapse involving such systems must lead to formation of singularities. On the other hand if there exists positive terms in the right hand side (e.g. $\nabla_{\alpha}a^{\alpha}$),
they offer repulsive contributions. If these terms happen to dominate, focusing will be prevented and the collapse will be averted. In the following sections we examine whether presence of magnetic fields offer repulsive contributions in dynamical situations and whether these contributions (if they exist) can overpower the converging ones.

\section{Homogeneous non-static spacetime}\label{sec3}
At first we briefly revisit the case of homogeneous non-static spacetimes. The utility of using the Raychaudhuri equation can be easily understood in this case. Due to presence of a magnetic field, the spacetime will be anisotropic. We consider a Bianchi I spacetime consisting of a fluid and a magnetic field of strength $B(t)$ along the $z$-direction in the rest frame of the fluid, 
\begin{equation}\label{metricih}
 \mathrm{d}s^2= -\mathrm{d}t^2+e^{2\lambda(t)}\left(\mathrm{d}x^2+\mathrm{d}y^2\right)+e^{2\gamma(t)}\mathrm{d}z^2.
\end{equation}

Now, for the comoving congruence $u^\alpha=\delta^\alpha_0$, we have $a^\alpha=0$ and $\omega_{\alpha\beta}=0$ i.e. the congruence consists of hypersurface orthogonal geodesics. 
We can decompose $\left(T_{\alpha\beta}u^\alpha u^\beta+\frac{1}{2}T\right)$ into two parts as, 
\begin{equation}\label{fmem}
 \left(T_{\alpha\beta}u^\alpha u^\beta+\frac{1}{2}T\right)=\left(T_{\alpha\beta}u^\alpha u^\beta+\frac{1}{2}T\right)_\mathrm{f}+\left(T_{\alpha\beta}u^\alpha u^\beta\right)_\mathrm{B}.
\end{equation}
The subscripts f, B respectively denote contribution from the fluid and the magnetic field hereafter. Here, we have used the fact that the energy-momentum tensor corresponding to the magnetic field has zero trace. $\left(T_{\alpha\beta}u^\alpha u^\beta\right)_\mathrm{B}$ is the energy density of the magnetic field as inferred by a comoving observer. Hence, this is always non-negative. Therefore, for a fluid satisfying the strong energy condition we must have $R_{\alpha\beta}u^\alpha u^\beta\geq 0$ by virtue of the Einstein equations. This implies irrespective of the presence of the magnetic field, focusing of the congruence within a finite time (in past or future) is inevitable. So, avoiding a singularity is not possible. Thorne considered all possible evolution scenarios of this model and found the corresponding solutions of the field equations\cite{thorne2}. All of them was found to incorporate singularities. We have arrived at the same conclusion using the Raychaudhuri equation and without referring to any exact solution of the field equations. So, we can see that the Raychaudhuri equation leads to important general conclusions even without any explicit solution of the field equations. This is a useful aspect of this Raychaudhuri equation based approach (for example see \cite{tsagas3,tsagasmr,sgc}).

For a homogeneous distribution the four acceleration vanishes. Thus, any agent which can oppose convergence of a congruence is absent in the Raychaudhuri equation \eqref{raych-eq} (assuming hypersurface orthogonal congruences for which the rotation tensor vanishes\cite{wald,poisson}). Thus, it is expected that these conclusions in the case of Bianchi I model should be valid for general homogeneous models. It is worthwhile to mention at this stage that in \cite{tsagasmr}, the authors studied magnetized collapse on a perturbed Bianchi I background. Perturbation at linear level allows to consider closed spatial sections. Then one can also have a non-vanishing four acceleration as first-order perturbations with reference to the  Bianchi I background. In this study the authors found that the fate of the evolution of this model actually  depends on the initial conditions of the problem.  In the next section we consider inhomogeneous models of gravitational collapse incorporating magnetic fields for further investigations.

\section{Inhomogeneous models of gravitational collapse}\label{sec4}
Here we consider an inhomogeneous gravitational collapse of a charged fluid distribution in the presence of a magnetic field. Magnetised collapse in an inhomogeneous environment still lacks a considerable amount of research. Germani and Tsagas\cite{tsagas1} studied collapse of an inhomogeneous dust distribution which is weakly magnetised using the Raychaudhuri equation. Considering the magnetic field to be relatively weak, the authors continued with the assumption of spherically symmetric Tolman-Bondi background metric. They found that, as the collapse proceeds, the spherical symmetry is severely distorted due to the presence of the magnetic field. 

In this work, we consider our system to be cylindrically symmetric, and the magnetic field is directed along the axis of symmetry. We analyse the system under the ideal--MHD approximation. This ensures a vanishing electric field in the frame of the fluid. Moreover, the magnetic field lines are frozen in the fluid\cite{priest,parker,mestel}. Study of gravitational collapse of cylindrically symmetric systems is reported in several places,  e.g. in \cite{thorne3,shapiro,eche,david,lete,nolan,wang,nn,hara} and references therein. However, such systems having inhomogeneity and incorporating magnetic fields have not been explored significantly.
  The metric for the spacetime under consideration can in general be written in the Einstein-Rosen form as\cite{david,kram},
\begin{equation}\label{metric}
\begin{split}
 \mathrm{d}s^2= e^{2\mu(r,t)-2\nu(r,t)}\left(-dt^2+dr^2\right)+e^{2\nu(r,t)}dz^2  +R^2(r,t)e^{-2\nu(r,t)}d\phi^2.
 \end{split}
 \end{equation}
 The strength of the magnetic field is $B(r,t)$ and its direction is along the $z$-axis in the rest frame of the fluid.
The energy-momentum tensor for the system is,
\begin{equation}
 T_{\alpha\beta}=\left(T_{\alpha\beta}\right)_\mathrm{f}+\left(T_{\alpha\beta}\right)_\mathrm{B},
\end{equation}
where
\begin{equation}
\begin{split}
 \left(T_{\alpha\beta}\right)_\mathrm{f}
=\left(\rho_\mathrm{f}+p_\mathrm{tf}\right)u_\alpha u_\beta+p_\mathrm{tf} g_{\alpha\beta}
+\left(p_\mathrm{rf}-p_\mathrm{tf}\right)\chi_\alpha \chi_\beta+ q_\mathrm{f}\left(u_\alpha \chi_\beta+ u_\beta \chi_\alpha\right),
\end{split}
\end{equation}
and\cite{barrow}
\begin{equation}
 \left(T_{\alpha\beta}\right)_\mathrm{B}=\left(\rho_\mathrm{B}+p_\mathrm{B}\right)u_\alpha u_\beta+p_\mathrm{B} g_{\alpha\beta}+\pi_{\alpha\beta}.
\end{equation}
Here $\rho_\mathrm{f}$, $p_\mathrm{rf}$, $p_\mathrm{tf}$ and $q_\mathrm{f}$ are the energy density, radial pressure, tangential pressure and radial heat flux of the fluid; $\rho_\mathrm{B}$ is the energy density of the magnetic field  given by, $\rho_\mathrm{B}=\frac{B^2}{2}$; $p_\mathrm{B}$ and $\pi_{\alpha\beta}$  are the  isotropic and anisotropic parts of the pressure of the magnetic field, $p_\mathrm{B}=\frac{B^2}{6}$ and $\pi_{\alpha\beta}=\frac{B^2}{3}h_{\alpha\beta}- B_{\alpha}B_{\beta}$; $B^\alpha=B n^\alpha$,  with $n^\alpha$ being a unit vector along $z$-direction; $\chi^\alpha$ is a unit vector along the radial direction and $u^\alpha$ is the velocity of the fluid. These unit vectors are respectively given by,
\begin{equation}\label{uvs}
 u^\alpha=e^{\nu-\mu}\delta^\alpha_0, \hspace{0.2cm} \chi^\alpha = e^{\nu-\mu}\delta^\alpha_1, \hspace{0.2cm} n^\alpha=e^{-\nu}\delta^\alpha_2, 
\end{equation}
where $u^\alpha u_\alpha=-1$, $u^\alpha\chi_\alpha=0$ and $u^\alpha n_\alpha=0$.

\subsection{Field equations}
For this system the Einstein field equations,
\begin{equation}
 G_{\alpha\beta}=\kappa T_{\alpha\beta},
\end{equation}
(where $G_{\alpha\beta}$ is the Einstein tensor) yield,

\begin{equation}\label{fe1}
\begin{split}
 G_{00}=\kappa T_{00} \implies  
  \frac{\dot{R}}{R}\dot{\mu}+\frac{R^\prime}{R}\mu^{\prime}-\left(\dot{\nu}^2+{\nu^\prime}^2\right)-\frac{R^{\prime\prime}}{R}=\kappa e^{2\mu-2\nu}\left(\rho_\mathrm{f}+\frac{B^2}{2}\right),
 \end{split}
 \end{equation}

\begin{equation}\label{fe2}
 G_{01}=\kappa T_{01}\implies
  \dot{\mu}\frac{R^\prime}{R}+\frac{\dot{R}}{R}\mu^\prime-2\dot{\nu}\nu^\prime-\frac{\dot{R}^\prime}{R}=-\kappa  e^{2\mu-2\nu} q_\mathrm{f},
 \end{equation}

\begin{equation}\label{fe3}
\begin{split}
 G_{11}=\kappa T_{11}\implies 
 \frac{\dot{R}}{R}\dot{\mu}-\frac{\ddot{R}}{R}+\frac{R^\prime}{R}\mu^{\prime}-\left(\dot{\nu}^2+{\nu^\prime}^2\right)=\kappa e^{2\mu-2\nu}\left(p_\mathrm{rf}+\frac{B^2}{2}\right),
 \end{split}
 \end{equation}

\begin{equation}\label{fe4}
\begin{split}
 G_{22}=\kappa T_{22}\implies 
  2\frac{\dot{R}}{R}\dot{\nu}-\frac{\ddot{R}}{R}-2\frac{R^\prime}{R}\nu^{\prime}+\frac{R^{\prime\prime}}{R}-\left(\dot{\nu}^2-{\nu^\prime}^2+\ddot{\mu}-2\ddot{\nu}-\mu^{\prime\prime}+2\nu^{\prime\prime}\right)=\kappa e^{2\mu-2\nu}\left(p_\mathrm{tf}-\frac{B^2}{2}\right),
 \end{split}
\end{equation}

\begin{equation}\label{fe5}
 G_{33}=\kappa T_{33}\implies 
  \mu^{\prime\prime}+{\nu^\prime}^2-\ddot{\mu}-\dot{\nu}^2=\kappa e^{2\mu-2\nu}\left(p_\mathrm{tf}+\frac{B^2}{2}\right),
 \end{equation}
where dot and prime in the superscript denote derivatives with respect to $t$ and $r$ respectively.

\subsection{Focusing Condition}
In this case, as the distribution contains charged particles in a magnetic field, the worldlines of these particles are not geodesics. This leads to a non-zero four acceleration of fluid particles.
For this model, the focusing condition,
\begin{equation}
 \frac{d\theta}{d\tau}+\frac{1}{3}\theta^2\leq 0 \implies R_{\alpha\beta} u^\alpha u^\beta+\sigma_{\alpha\beta}\sigma^{\alpha\beta}-\nabla_\alpha a^\alpha \geq 0.
\end{equation}
We have used the Raychaudhuri equation \eqref{raych-eq} in the above expression.
The divergence of acceleration term  has been shown to be related to the magneto-curvature stress in \cite{tsagas2,tsagas4}. This stress acts against the gravitational pull. Therefore it resists focusing of particle worldlines and the ultimate collapse to a singularity.  We will now investigate whether the divergence of acceleration ($\nabla_\alpha a^\alpha$) can balance or even dominate over the combined effect of the curvature ($R_{\alpha\beta} u^\alpha u^\beta$) and shear ($\sigma_{\alpha\beta}\sigma^{\alpha\beta}$).  

It is difficult to draw useful conclusions in a completely general setting. Therefore, we need to consider simplified models in order to understand the explicit role of the magnetic field towards the dynamics of collapse of inhomogeneous models. We will consider two special cases - in the first case we assume that the metric coefficients are separable in spatial and temporal dependent parts whereas we assume the spacetime to be self-similar in the second one.

\subsection{Case I}
As the first example we consider an ansatz,
\begin{equation}\label{ansatz1}
 \mu=k \nu, \hspace{0.2cm} \nu(r,t)=\nu_1(t)+\nu_2(r) \hspace{0.2cm} \mathrm{and} \hspace{0.2cm} R^2(r,t)=r^2 e^{2m\nu_1(t)}, 
\end{equation}
where $k, m$ are constants.
With this ansatz, we can express all the geometric quantities in the Raychaudhuri equation in terms of the matter variables. Manipulation of the field equations \eqref{fe1}-\eqref{fe5} in this case yields,
\begin{equation}\label{rmns}
 R_{\alpha\beta} u^\alpha u^\beta=\frac{\rho_\mathrm{f}+p_\mathrm{rf}+2 p_\mathrm{tf}}{2}+\frac{B^2}{2},
\end{equation}

\begin{equation}\label{divaccs}
 \nabla_\alpha a^\alpha=\kappa \left[\frac{k-1}{2}B^2+(k-1)\left(\frac{1}{m}-\frac{1}{2}\right)\left(\rho_\mathrm{f}-p_\mathrm{rf}\right)\right],
\end{equation}
 and
\begin{equation}\label{sigs}
 \sigma_{\alpha\beta}\sigma^{\alpha\beta}=\frac{2}{3}e^{2(1-k)\nu}\left(m^2+k^2-m(k+2)-2k+4\right)\dot{\nu_1}^2,
\end{equation}
where 
\begin{equation}\label{nu1dots}
\begin{split}
 \dot{\nu_1}^2 =\kappa e^{2(k-1)\nu}\left[\frac{2-k}{4(mk-1)}B^2+\frac{k+2}{4(mk-1)}\left(\rho_\mathrm{f}-p_\mathrm{rf}\right)\right. +\left. \frac{p_\mathrm{rf}+p_\mathrm{tf}}{2(mk-1)}\right].
 \end{split}
 \end{equation}
The primarily reason behind invoking such a stringent requirement is that it enables us to extract useful information about the system through an analytical study using the adopted approach. Nevertheless, this scenario is not completely unphysical. The matter distribution is quite general, we have not assumed anything about the strength of the magnetic field at the outset and the evolution is still inhomogeneous.

For the violation of the focusing condition we need the quantity, $R_{\alpha\beta} u^\alpha u^\beta-\nabla_\alpha a^\alpha+\sigma_{\alpha\beta}\sigma^{\alpha\beta}$ to be negative.  Here the constants $k$ and $m$ play a crucial role in deciding the structure of the spacetime. If for certain values of $k$ and $m$  the divergence of acceleration dominates at some stage of the evolution, collapse will halt, and the evolution may even undergo an expansion. Otherwise, collapse continues towards a singularity. 

Let us consider the case of $m=2$ for illustration. In this case the divergence of acceleration contains contribution from the magnetic field only. This is given by,
\begin{equation}
 \nabla_\alpha a^\alpha= \frac{\kappa(k-1)}{2}B^2.
\end{equation}
Therefore, the magnetic field offers an effect opposing gravity if $k>1$.
Here we have,
\begin{equation}\label{fcs}
\begin{split}
  R_{\alpha\beta} u^\alpha u^\beta+\sigma_{\alpha\beta}\sigma^{\alpha\beta}-\nabla_\alpha a^\alpha =  \kappa\left[\frac{(k+1)^2(2-k)}{6(2k-1)}B^2\right.  \left.+\frac{(k+1)(k^2-3k+5)}{6(2k-1)}(\rho_\mathrm{f}-p_\mathrm{rf})+\frac{4k-1}{2(2k-1)}(p_\mathrm{rf}+p_\mathrm{tf})\right].
 \end{split}
\end{equation}
When the magnetic field is switched off the divergence of acceleration vanishes and we have $R_{\alpha\beta} u^\alpha u^\beta+\sigma_{\alpha\beta}\sigma^{\alpha\beta}\geq 0$ if the strong energy condition holds. Therefore in the right hand side of expression \eqref{fcs}, the first term which is related to the  magnetic field strength must give a negative contribution and dominate over the other two terms so that the focusing condition is violated (here we should note that there is an interaction present between the magnetic field and the charged fluid, but it is reasonable to assume that when $B=0$ other two terms in the right hand side of equation \eqref{fcs} give a positive contribution). For having this, we must have $k>2$ (as we have to incorporate $k>1$ as already mentioned) and 
\begin{equation}\label{icond1}
 B^2> \frac{k(k-3)+5}{(k+1)(k-2)}\left(\rho_\mathrm{f}-p_\mathrm{rf}\right)+\frac{3(4k-1)}{(k+1)^2(k-2)}\left(p_\mathrm{rf}+p_\mathrm{tf}\right).
\end{equation}
We can infer that if the magnetic field is sufficiently large to overcome the gravitating effect of the fluid so that the condition \eqref{icond1} holds, focusing of the congruence can be prevented. But, the magnetic field cannot be arbitrarily large. We have an upper bound on the strength of the magnetic field. This comes form the condition that $\dot{\nu_1}^2$ given by equation \eqref{nu1dots} must always be positive,
\begin{equation}
 \dot{\nu_1}^2> 0 \implies B^2< \frac{k+2}{k-2}\left(\rho_\mathrm{f}-p_\mathrm{rf}\right)+\frac{2\left(p_\mathrm{rf}+p_\mathrm{tf}\right)}{(k-2)}.
\end{equation}
Therefore, to avoid focusing and hence the final singularity formation due to a continuous collapse, the magnetic field strength should satisfy the following constraint,
\begin{equation}\label{ineq1}
\begin{split}
 \left[\frac{k(k-3)+5}{(k+1)(k-2)}\left(\rho_\mathrm{f}-p_\mathrm{rf}\right)+\frac{3(4k-1)}{(k+1)^2(k-2)}\left(p_\mathrm{rf}+p_\mathrm{tf}\right)\right] <B^2< \left[\frac{k+2}{k-2}\left(\rho_\mathrm{f}-p_\mathrm{rf}\right)+\frac{2\left(p_\mathrm{rf}+p_\mathrm{tf}\right)}{(k-2)}\right].
 \end{split}
\end{equation}
For conventional fluids we have $\rho_\mathrm{f}\geq p_\mathrm{rf}$ and $p_\mathrm{rf}, p_\mathrm{tf}\geq 0$. Therefore, the upper and lower bounds of the above inequality \eqref{ineq1} are always positive (as we continue with the assumption that $k>2$). One can also note that the upper bound is manifestly greater than the lower bound when $k>\frac{4+\sqrt{6}}{2}$.

At this stage it should be noted that if we choose   $k=2$ and $m=2$, this model leads to a simple conformal collapse scenario. The metric \eqref{metric} is then given by, 
\begin{equation}\label{metricconf}
\begin{split}
 \mathrm{d}s^2= e^{2\nu_1(t)}\left[e^{2\nu_2(r)}\left(-dt^2+dr^2+dz^2\right) \right. \left. +r^2 e^{-2\nu_2(r)}d\phi^2\right].
 \end{split}
 \end{equation}
Conformal collapse models are gaining increasing interest in recent times\cite{chakra,bini}. When we choose $e^{2\nu_2} = (1+r^2)^2$, then the conformally related metric (within the third bracket in equation \eqref{metricconf}) is the well-known Melvin static metric. Bini and Mashhoon\cite{bini} considered this kind of a dynamical spacetime represented by equation \eqref{metricconf} with $e^{2\nu_2} = (1+r^2)^2$ and $e^{2\nu_1}=(a+bt)$. The authors then found the corresponding energy-momentum tensor. Their aim was to consider a gravitational collapse where the seed is the Melvin spacetime. This is a special case of our model characterized by the ansatz \eqref{ansatz1}. In our case, it is evident from \eqref{fcs} that formation of singularity cannot be avoided when $k=2$ if we have conventional fluids.

\subsection{Case II}
Next, we consider a self-similar spacetime of the first kind. This spacetime is characterized by the presence of a homothetic Killing vector\cite{maar,moop1,moop2}. Self-similarity has been found to play an important role in the description of the asymptotic nature of different models in Astrophysics and Cosmology. Collapse of a star towards a singularity, evolution of the universe from the Big Bang etc. tend to follow self-similar nature in some situations. Another important context where self-similarity features itself is in the formation of Naked singularities. 
One of the most important fields of application of self-similar solutions is the Critical Phenomena in gravitational collapse. This is one of the most exciting discoveries of the recent years in the field of GR. Usually, the critical phenomena is studied using numerical methods and found to involve discrete self-similarity. Discrete case is difficult to handle in analytical studies. Therefore, continuous self-similarity is assumed to carry out such studies. 
For a comprehensive discussion regarding the importance of self-similarity in gravitational physics and the above mentioned points we refer to \cite{car2,car} and references therein.
 An elegant discussion regarding the critical phenomena can be found in the review \cite{Gundlach1}.  
 
 Cylindrically symmetric models involving self-similar solutions have also been studied.
 In a numerical study it has been shown that the asymptotic nature of the exterior gravitational field of a collapsing hollow dust cylinder is given by self-similar cylindrically symmetric vacuum solutions\cite{naka}. Self-similar collapse in cylindrical symmetry has been investigated in the works \cite{hara,wang1,sharif,lvn,condron}. Behavior of the kinematical quantities like expansion, shear, rotation etc. for self-similar cylindrically symmetric solutions was studied by Sharif and Aziz\cite{sharif1}. They have also discussed the singularity feature of these solutions. \\

With the assumption of a continuous self-similarity, the cylindrically symmetric metric \eqref{metric} can be written as\cite{lvn},
\begin{equation}
\begin{split}
  \mathrm{d}s^2 = e^{2\mu(\xi)-2\nu(\xi)}\left(-dt^2+dr^2\right)+ r^2 \left[e^{2\nu(\xi)}dz^2 \right. \left. +R^2(\xi)e^{-2\nu(\xi)}d\phi^2\right],
  \end{split}
  \end{equation}
using a redefinition of the metric coefficients and the coordinates. Here $\xi$ is the self-similar variable given by, $\xi=\frac{t}{r}$. We use a similar simplification as in the previous case by considering the following ansatz,
\begin{equation}
 \mu=k\nu \hspace{0.2cm} \mathrm{and} \hspace{0.2cm} R=e^{m \nu}.
\end{equation}
Here we have, $u^\alpha=e^{\nu-\mu}\delta^\alpha_0$,  $\chi^\alpha = e^{\nu-\mu}\delta^\alpha_1$ and $ n^\alpha=\frac{e^{-\nu}}{r}\delta^\alpha_2$.
From Einstein's field equations it follows that,
\begin{equation}\label{ricss}
 R_{\alpha\beta} u^\alpha u^\beta=\frac{\rho_\mathrm{f}+p_\mathrm{rf}+2 p_\mathrm{tf}}{2}+\frac{B^2}{2},
\end{equation}
\begin{equation}\label{divaccss}
 \nabla_\alpha a^\alpha=  \frac{\kappa \xi^2\left(k-1\right)}{\left(\xi^2-1\right)\left(2-m\right)}B^2,
\end{equation}
and 
\begin{equation}\label{sigss}
 \sigma_{\alpha\beta}\sigma^{\alpha\beta} =\frac{2}{3}\frac{e^{2(1-k)\nu}}{r^2}\left(m^2+k^2-m(k+2)-2k+4\right)\tilde{\nu}^2,
 \end{equation}
where
\begin{equation}
\begin{split}
 \tilde{\nu}^2\equiv \left(\frac{\mathrm{d\nu}}{\mathrm{d}\xi}\right)^2=\frac{\kappa r^2 e^{2(k-1)\nu}}{\xi^2-1}\left[\frac{4-4k-m}{2(m-2)(km-1)}B^2\right.  +\frac{k-1}{m(km-1)}(\rho_\mathrm{f}-p_\mathrm{rf}+2)-\left. \frac{p_\mathrm{tf}}{km-1}\right].
 \end{split}
\end{equation}

In this setting, when $m=2$, we must have $B=0$ for consistency. So, the case of $m=2$ is not relevant in the current context. Following the similar arguments as discussed in the previous case, here also we can obtain a constraint (similar to \eqref{ineq1}) on the magnetic field strength for avoiding focusing of the congruence of particle worldlines. This will follow from the two conditions which are $R_{\alpha\beta} u^\alpha u^\beta+\sigma_{\alpha\beta}\sigma^{\alpha\beta}-\nabla_\alpha a^\alpha< 0$ and $\tilde{\nu}^2>0$. Here, it is to be noted that contrary to the previous case, fate of the focusing condition depends explicitly on the spacetime point (through the presence of $\xi$ in $\nabla_\alpha a^\alpha$ and $\tilde{\nu}^2$). 

We will consider two specific limits in this case. The first one is $\xi \rightarrow \infty$ which will give us an idea about the formation of a central singularity (another limit, $t\rightarrow \infty$ is inconsequential). It is clear from equations \eqref{ricss}, \eqref{divaccss} and \eqref{sigss} that $R_{\alpha\beta} u^\alpha u^\beta$, $\nabla_\alpha a^\alpha\gg \sigma_{\alpha\beta}\sigma^{\alpha\beta}$ in this limit and we can ignore the effect of shear. Thus, to avoid formation of a central singularity due to collapse we need,
\begin{equation}
\begin{split}
  R_{\alpha\beta} u^\alpha u^\beta-\nabla_\alpha a^\alpha<0\implies   \left[\frac{\left(k-1\right)}{\left(2-m\right)}-\frac{1}{2}\right]B^2>\frac{\left(\rho_\mathrm{f}+p_\mathrm{rf}+2p_\mathrm{tf}\right)}{2},
 \end{split}
 \end{equation}
and the constants $k$ and $m$ have to satisfy  $\frac{\left(k-1\right)}{\left(2-m\right)}>\frac{1}{2}$.

Another limit is when $t\rightarrow 0$, which corresponds to $\xi\rightarrow 0$. In this limit, $\nabla_\alpha a^\alpha\ll R_{\alpha\beta} u^\alpha u^\beta, \sigma_{\alpha\beta}\sigma^{\alpha\beta}$. So, the focusing condition will be obeyed. Now, there are two possibilities. The first one is that there will be a singularity at $t=0$. Alternatively, we can choose the initial conditions in such a way that the evolution starts from a contracting phase. As the evolution proceeds it ends up in a singularity formation if the focusing condition continues to hold. If on the other hand, $\nabla_\alpha a^\alpha$ becomes effective and starts to dominate during the evolution, the collapse can be halted, even an expansion may occur. This is quite possible because as the collapse proceeds magnetic field lines will be distorted more and more and the opposing effect of the magnetic field towards the contraction will thus increase. A quite intuitive and detailed discussion regarding this point can be found in the work of Tsagas and Mavrogiannis\cite{tsagas4}.\\

The examples discussed above are very simple and contain only one independent metric coefficient. But, they are useful enough to provide valuable insights regarding the role of the magnetic field in the collapse of inhomogeneous distributions containing charged matter.

\section{Conclusion}\label{sec5}
It has been found in many studies that Electromagnetism can play a significant role in gravitational collapse of a matter distribution and its effect can  alter the conventional final fate of a collapse. Considering magnetic fields in collapse scenarios is certainly a well-motivated choice as discussed in various works in the literature.
In this work we have explored  whether magnetic fields give rise to agents that can oppose gravity in inhomogeneous evolving spacetimes. Also, we have examined if there exists situations where such repulsive contributions can balance or overpower the contracting stresses. We have adopted an approach based on the Raychaudhuri equation for this study.

We have started with a brief review of the case of a homogeneous non-static distribution to discuss the utility of the Raychaudhuri equation based approach. In this case, we have found that any repulsive contribution does not arise due to the effect of the magnetic field. Hence, formation of singularities (in past or future) is inevitable. This conclusion is consistent with the existing literature and follows quite easily from the Raychaudhuri equation.

We have then gone on to study inhomogeneous collapse of a charged matter distribution in the presence of a magnetic field and within the ideal--MHD limit. The system is assumed to be cylindrically symmetric with the magnetic field directed along the axis of symmetry in the rest frame of the fluid. It is very difficult to draw  fruitful inferences for the completely general case. So, we have considered two special cases, namely the separable case and the self-similar case. In both of these scenarios we have found that the magnetic field provides repulsive effects. We have found constraints on the magnetic field strength for which focusing of worldlines of charged particles can be prevented and hence collapse can be averted. 

Our work strengthens the claims of \cite{tsagas2,tsagas3,tsagas4}.  The current work acquires importance since it allows us to draw important conclusions about inhomogeneous collapsing systems with magnetic fields in spite of the fact that no exact solution for such systems seems to be available in the literature.

To conclude with certainty that presence of a magnetic field prevents formation of a singularity, we do need solutions of the field equations. This will also determine what values of the parameters $k$ and $m$ are allowed and whether the constraints obtained on the magnetic field strength to avoid a singularity formation due to collapse are obeyed. A detailed study of this kind goes beyond the focus in the current paper. Our main idea in this work is to investigate the role of a magnetic field in deciding the structure of a collapsing spacetime through the study of evolution of timelike congruences. The Raychaudhuri equation enables us to carry out such a study without solving the field equations explicitly.

\section*{Acknowledgments}
The author acknowledges Council of Scientific and Industrial Research (CSIR), India for providing the financial support through a senior research fellowship (Award No. 09/921(0188)/2017-EMR-I). The author is thankful to Dr. Ananda Dasgupta and Prof. Narayan Banerjee for providing valuable insights and suggestions which improved the quality of the work. The author thanks Dr. Soumya Chakrabarti for useful discussions and comments. 

\section*{Data Availability Statement} The author neither
used nor generated any data while performing this work.

\bibliographystyle{elsarticle-num}

\end{document}